\documentclass[twocolumn,showpacs,superscriptaddress,prl]{revtex4}
\usepackage{amsbsy,amssymb,amsmath,bm,graphicx}
\newcommand{\da}{\downarrow}
\newcommand{\ua}{\uparrow}
\newcommand{\im}{i}

\newcommand{\bs}{\mathbf {s}}

\newcommand{\bS}{\mathbf {S}}

\newcommand{\cH}{\mathcal {H}}

\usepackage{color}
\newcommand{\ignore}[1]{}
\newcommand{\jComment}[1]{}
\newcommand{\yComment}[1]{}
 \renewcommand{\jComment}[1]{\textcolor{blue}{\sf  Joakim: #1}}
 \renewcommand{\yComment}[1]{\textcolor{green}{\sf Yuri: #1}}

\begin{document}
\title{Rabi oscillations of a qubit coupled to a two-level system}
\author{Y. M. Galperin}
\email{iouri.galperine@fys.uio.no}
\affiliation{Department of Physics, University of Oslo, PO Box 1048
  Blindern, 0316 Oslo, Norway}
\affiliation{A. F. Ioffe  Physico-Technical Institute of Russian Academy of
Sciences, 194021 St. Petersburg, Russia}
\affiliation{Argonne National Laboratory, 9700 S. Cass av., IL 60439, USA}
\author{D. V. Shantsev}

\affiliation{Department of Physics, University of Oslo, PO Box 1048
  Blindern, 0316 Oslo, Norway}
\affiliation{A. F. Ioffe  Physico-Technical Institute of Russian Academy of
Sciences, 194021 St. Petersburg, Russia}

\author{J. Bergli}
\affiliation{Physics Department, Princeton University,
        Princeton, NJ 08544, USA}
\author{B. L. Altshuler}
\affiliation{Physics Department, Princeton University,
        Princeton, NJ 08544, USA}
\affiliation{NEC-Laboratories America, Inc., 4 Independence Way, Princeton, NJ 08540, USA}

\date{ Received 
}

\begin{abstract}
The problem of Rabi oscillations in a qubit coupled to a fluctuator
and in contact with a heath bath is considered. A scheme is developed
for taking into account both phase and energy relaxation in a
phenomenological way, while taking full account of the quantum
dynamics of the four-level system subject to a driving AC
field. Significant suppression of the Rabi oscillations is found when
the qubit and fluctuator are close to resonance. The
effect of the fluctuator state on the read-out signal is
discussed. This effect is shown to modify the observed signal
significantly. This may be relevant to recent experiments by Simmonds
{\it et al.} [Phys. Rev. Lett. {\bf 93}, 077003 (2004)].

\end{abstract}


\maketitle

Recent experiments have demonstrated Rabi oscillations in macroscopic
quantum systems
(qubits).~\cite{Martinis,Martinis1,Nakamura,Martinis2,Vion,Yu,Nakamura2}
These oscillations decay rather quickly due to interaction with
the environment even in pure systems. An interesting behavior was
observed in spectroscopic experiments with Josephson
qubits.~\cite{Martinis} Both the decay rate and the oscillation
pattern were strongly dependent on the qubit eigenfrequency ramped by
external parameters. Namely the oscillations were significantly
suppressed in the vicinity of certain eigenfrequencies. This was
interpreted~\cite{Martinis} as an influence of some two-level systems
(fluctuators) located in the qubit environment. The suppression is
strong when the fluctuator's and qubit's energy splittings are very
close.

 Below we present a theory of a qubit interacting with an external AC
 field and a single fluctuator coupled to the qubit. In addition, we
 assume that the system interacts with some thermal bath providing
 both phase and energy relaxation.
The account of the energy relaxation distinguishes our model from a
similar approach used in Ref.~\onlinecite{Loss}, whereas in the recent 
numerical work of Ref.~\onlinecite{Ku} no account is taken of the 
phase relaxation.  
By solving the kinetic equation for the density matrix we compute level
 populations versus time. The results explain the strong influence of the
 resonant fluctuator on the Rabi oscillations of the qubit in
 agreement with experimental findings.

We will characterize the qubit by a spin
$\bS$ interacting with a fluctuator represented by a spin
$\bs$. The
Hamiltonian in the form
\begin{equation}
  \label{eq:001}
  \tilde{\cH}(t)=\cH_{\text{q}}+\cH_{\text{f}}+\cH_{\text{q-f}}
+\cH_{\text{man}}(t)\,. 
\end{equation}
takes into account the qubit and the fluctuator ($\cH_{\text{q}}$ and 
$\cH_{\text{f}}$) as well as the interaction between them ($\cH_{\text{q-f}}$).
We also included into $\tilde{\cH}(t)$ the 
``manipulation'' part $ \cH_{\text{man}}(t) =S_x
F\cos\omega t$, which in spin terms is an oscillating magnetic field in the 
$x$-direction applied to the qubit ($F$ is the Rabi frequency). 
The other parts of the Hamiltonian can 
be written as
\begin{eqnarray}
  \label{eq:002}
&& \cH_{\text{q}} = \frac{E}{2}S_z, \   \cH_{\text{f}}=\frac{e}{2}s_z,
  \cH_{\text{q-f}}=
\frac{u}{2}(S_xs_x+S_ys_y)\, .
\end{eqnarray}
Here $E$ is the distance between the qubit
levels, $e$ is the distance 
between the fluctuator's levels, $S_i$ and $s_i$ are the Pauli
matrices acting respectively in the spaces of the qubit and
fluctuator, $u$ is the off-diagonal coupling constant.
\footnote{
We have taken into account only the off-diagonal coupling, $(u/2)({\bf
  S}_\perp \cdot {\bf s}_\perp)$, 
between the qubit and fluctuator. One can also include
$\cH_z=(v/2)({\bf S}_z \cdot {\bf s}_z)$. 
This interaction contributes to the decoherence of the
qubit via random modulation of the Rabi frequency.~\cite{GAS}
However, as we checked, it does not specifically affect the visibility of Rabi
oscillations as long as the fluctuator is decoupled from the AC field,
$\cH_{12}=\cH_{34}= 0$. 
}
Below we will
use the so-called resonant approximation replacing
$\cos \omega t \to (1/2)\exp(i\omega t)$, see {\it e.g} Slichter\cite{book}.
This approximation, which neglects
higher harmonics of the response, is valid
close to the resonance, i. e., when 
$|E-\omega|,|e-\omega|,  |u| \ll \omega$.
We also omit the diagonal
qubit-fluctuator interaction $\propto S_zs_z$ which is important only when 
the qubit and the fluctuator are far from
the resonance. On the contrary, the
off-diagonal part is important only when the qubit and the fluctuator
have close energy  splitting, $|E-e| \ll E,e$. The importance of this
interaction was stressed in Ref.~\onlinecite{Martinis}. 
We will also assume that
\begin{equation}
  \label{eq:003}
  |u| \ll \omega \lesssim T\, ,
\end{equation}
where $\omega$ and $T$ are measured in energy units.
Only in this case the qubit acts as a resonant system. 

We are not going to take the thermal bath explicitly into
account. Instead we introduce damping
phenomenologically into the equation for the density matrix evolution.  

The Hamiltonian (\ref{eq:001}) in the resonant approximation
can be expressed as a $4\times4$ matrix 
\begin{equation}
  \label{eq:04a}
  \cH = \frac{1}{2}\, \left(\begin{array}{cccc}
-E-e&0&Fe^{\im \omega t}&0\\
0&-E+e & u &Fe^{\im \omega t}\\
Fe^{-\im\omega t}&u^*&E-e&0\\
0&Fe^{-\im \omega t}&0&E+e
\end{array} \right)\, .
\end{equation}
\begin{figure}[h]
\centerline{
\includegraphics[width=6cm] {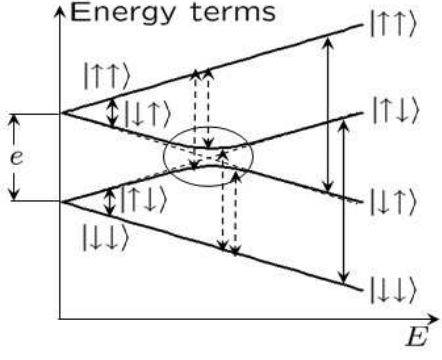}}
\caption{Diagram of the terms of the 4-level system
  consisting of the qubit and the fluctuator. Shown by arrows are transitions
  induced by the AC field. 
\label{fig:01}} 
\end{figure}
Shown in Fig.~\ref{fig:01} are the energy terms of the Hamiltonian  
versus the qubit energy splitting $E$ for a fixed fluctuator splitting
$e$ and $u=0.1 e$. Near the resonance, $E=e$, all 4 levels are
involved in the AC-induced transitions. In this region one can expect
strong influence of the fluctuator on the qubit response. 

The 4-level system qubit+fluctuator is characterized by a 4$\times$4 density 
matrix $\tilde{\rho}_{\mu\nu}(t)$ which diagonal matrix elements 
$n_{\da \da},n_{\da \ua}, n_{\ua \da}, n_{\ua \ua}$ describe the 
occupations of each of the levels. In the presence of the high-frequency
driving force $\cos\omega t$ the off-diagonal matrix elements also quickly 
oscillate:
\begin{equation}
  \label{eq:05a}
 \left( \begin{array}{cccc}
n_{\da \da}& -\im f\, e^{\im \omega t}& -\im g\, e^{\im \omega t}&
-\im j\, e^{2 \im \omega t}\\  
\im f^*\, e^{-\im \omega t} &n_{\da \ua}& -\im k& -\im l\, e^{\im \omega t}\\
\im g^*\, e^{-\im \omega t}& \im k^*& n_{\ua \da}&- \im m\, e^{\im \omega t}\\
\im j^*\, e^{-2\im \omega t}& \im g^*\, e^{-\im \omega t}& \im m^*\,
e^{-\im \omega t}&n_{\ua \ua}     
\end{array}\right)\,.
\end{equation}
It is convenient to introduce a matrix with elements that all vary slowly 
with time:
\begin{equation}
  \label{eq:05}
  \hat{\rho}_{\mu\nu}\equiv\left( \begin{array}{cccc}
n_{\da \da}& -\im f& -\im g& -\im j\\ 
\im f^* &n_{\da \ua}& -\im k& -\im l\\
\im g^*& \im k^*& n_{\ua \da}&- \im m\\
\im j^*& \im l^*& \im m^*&n_{\ua \ua}   
\end{array}\right)
\end{equation}
The matrix (\ref{eq:05}) is just the density matrix in the
 frame rotating with frequency $\omega$ around the $z$-axis in the qubit spin
space. In this
frame the von Neumann equation,  $\partial_t\hat{\rho}=\im
[\cH,\hat{\rho}]$, involves a time-independent Hamiltonian, 
\begin{equation}
  \label{eq:04}
  \cH = \frac{F}{2}\, \left(\begin{array}{cccc}
-{\cal E}-\epsilon&0&1&0\\
0&-{\cal E}+\epsilon & \eta &1\\
1&\eta^*&{\cal E}-\epsilon&0\\
0&1&0&{\cal E}+\epsilon
\end{array} \right)\, .
\end{equation}
Here ${\cal E}=(E-\omega)/F, \, \epsilon=(e-\omega)/F, \,  \eta=u/F$. The
matrix elements $\cH_{13}=\cH_{31}^*$ and $\cH_{24}=\cH_{42}^*$
correspond to transitions between the qubit levels, while
$\cH_{23}=\cH_{32}^*$ describe simultaneous ``flips'' of the
qubit and the fluctuator. 

Our next step is to rewrite the von Neumann equation
\begin{equation} \label{meq1}
\dot{\bm{\rho}}=\hat{L}\, \bm{\rho}
\end{equation}
as an equation 
for the vector $\bm{\rho}$ with 16 components
\begin{equation} \label{v01}
\{n_{\da \da},n_{\da \ua}, n_{\ua
  \da}, n_{\ua \ua},g
  ,g^*\!,l,l^*\!,f,f^*\!,m,m^*\!,
  k,k^*\!,j,j^* \}.
\end{equation}
The 16$\times$16  evolution matrix $\hat{L}$
is read off 
from the commutator $[\cH,\rho]$.

When no heat bath is coupled to the 4-level system, 
the matrix $\hat{L}$ has rank 12. This
 reflects the fact that in the
absence of dissipation there are four conserved quantities, namely the 
occupation numbers of the eigenstates of the Hamiltonian Eq. (\ref{eq:04}).

The system  reaches a stationary
state only in the presence of dissipation.
We take the dissipation into account in the following way. 
First, we assume 
that at large time the system will reach a stationary, however
non-equilibrium, state $\bm{\rho}^\infty$  where the matrix elements
depend on the 
relationship between the pumping amplitude, $F$, and the phase and
energy relaxation rates. 
Given the ``dissipation operator'' $\hat{\Gamma}$, which takes into
account both phase and energy relaxation one can determine  $\bm{\rho}^\infty$
as a solution of the equation
\begin{equation}
  \label{eq:dmst01}
  (\hat{L}+\hat{\Gamma})\bm{\rho}^\infty = 0\, .
\end{equation}

The explicit form of the dissipation operator  $\hat{\Gamma}$ is determined 
by the particular model chosen for the thermal bath. At the same time, the 
diagonal matrix elements of the stationary density matrix 
\begin{equation}
  \label{eq:v02}
  \bm{\rho}^\infty_{\text{diag}}=\{n_{\da \da}^\infty,n_{\da
  \ua}^\infty, n_{\ua 
  \da}^\infty, n_{\ua \ua}^\infty\}\, ,
\end{equation}
can be specified phenomenologically in a model-independent way. The dynamics 
of the remaining 12 matrix elements $\bm{\rho}^\infty_{\text{off-diag}}$ is 
rather insensitive to the details of the model. 

The situation here is 
somewhat similar to nonlinear AC conductivity in disordered metals: 
when the scattering of itinerant electrons is predominantly elastic, 
the details of the inelastic relaxation are not important in spite of the 
Joule heat. This inelastic relaxation rate enters usually only though
the effective field-dependent temperature, while the 
temperature-dependent conductivity can be determined without taking 
energy relaxation into account. 

Adopting this strategy we choose an approximate dissipation operator 
(16$\times$16 matrix) $\hat{\Gamma}_0$ and present the operator 
$\hat{L}+\hat{\Gamma}_0$ as 

\begin{equation}
  \hat{L}+\hat{\Gamma}_0 
  = \begin{array}{cc}\left(
   \begin{array}{c|c}\hat{M}_2&\hat{M}_1^\dag\\
    \hline
    \hat{M}_1&\hat{M}_0\end{array}
    \right) & \begin{array}{c}4\\12\end{array} \\
    \begin{array}{cc}4&12\end{array}&
    \end{array}
\end{equation}

The off-diagonal and diagonal elements of the density matrix $\bm{\rho}^\infty$
are connected through the 12$\times$12 matrix $\hat{M}_0$ and the 12$\times$4
matrix $\hat{M}_1$:
\begin{equation}
  \label{eq:rhoinfty}
  \bm{\rho}^\infty_{\text{off-diag}} =-\hat{M}^{-1}_0\hat{M}_1
  \bm{\rho}_\text{diag}^\infty \, .
\end{equation}

The advantage of this method is that as soon as stationary level
populations are specified the results only weakly depend on the
specific form of the matrix $\hat{\Gamma}_0$. For simplicity  
we adopt a model similar to the one conventionally used to
describe relaxation in various two-level systems, see, e. g.,
Ref.~\onlinecite{book}. If the qubit does not directly interact with
the environment, while the fluctuator has a finite dephasing rate
$\gamma$ and energy relaxation rate $2\gamma$, then $\hat{\Gamma}_0$ 
is diagonal $(\hat{\Gamma}_0)_{ik} = \gamma_i\delta_{ik}$
where
$$\gamma_i= \left\{\begin{array}{rl}
-2\gamma\, ,& 1\le  i \le 4 \\ 
- \gamma\, ,& 9\le  i \le 12 \, .
\end{array}\right. $$
This damping matrix differs from the one used in Ref.~\onlinecite{Loss} by 
taking into account 
relaxation of the diagonal elements of the
density matrix.

Given $\bm{\rho}^\infty$,  the solution of the equation 
$\dot{\bm{\rho}} (t)= (\hat{L}+\hat{\Gamma}_0)\bm{\rho}(t)$
can be presented  as an expansion 
\begin{equation}
  \label{eq:07}
  \bm{\rho}(t)=\bm{\rho}^\infty +\sum_{i=2}^{16} c_i \bm{e}_i e^{\lambda_i t}  
\end{equation}
over the eigenvectors, $\bm{e}_i$,  of the matrix $\hat{L}+\hat{\Gamma}_0$, 
$\lambda_i$ being the corresponding eigenvalues. For weak dissipation 
$\bm{e}_i$ and $\lambda_i$ are close correspondingly to 
$\bm{e}_i^{(0)}$ and $\lambda_i^{(0)}$ - eigenvectors and eigenvalues of 
$\hat{L}$. We start summation in Eq. (\ref{eq:07}) from $i=2$ to 
emphasize that $\lambda_1=0$ regardless of the dissipation if we choose 
$\bm{e}_1$ such that its scalar product with $\bm{\rho}$ is $tr \hat{\rho}
= n_{\da \da}+n_{\da \ua}+ n_{\ua\da}+ n_{\ua \ua}$. Accordingly, the part of 
$\bm{\rho}(t)$ proportional to $\bm{e}_1$ should be included into 
$\bm{\rho}^\infty$. Note that in general $\bm{\rho}^\infty$ can not be 
presented as a linear combination only of $\bm{e}_i$ with $i\leq4$.

One can specify the initial conditions 
assuming
that the external
AC field is switched on at $t=0$:
\begin{equation}
  \label{eq:08}
  \rho_{11} \equiv n_{\da \da} (0) =1, \ \rho_{i\ne 1,k\ne 1}(0)=0\, .  
\end{equation}
We have checked numerically 
that the results are only weakly sensitive to initial
conditions.
It is more important to determine $\bm{\rho}^\infty$, which 
strongly depends on the pumping strength, $F$, as
well as on the relaxation times of the phase, $\tau_2$,  and energy, $\tau_1$.
For strong pumping $F^2 \tau_1 \tau_2 \gg 1$
the levels with different orientations of the qubit have almost the same
occupations. The stationary populations of the fluctuator level depend
on the interplay between its coupling with the qubit
and the energy relaxation rate. If the coupling
is weak, the stationary occupancies of the levels
are given by a thermal distribution:
$$n_{\ua \ua}=n_{\da \ua} \, , \quad   n_{\ua \da}=n_{\da \da} \, ,
  \quad  n_{\ua
  \ua}/n_{\ua \da}=e^{-e/kT}\, . $$
Since we do not focus on the temperature dependence we will 
set the ratio 
$  n_{\ua \ua}/n_{\ua \da} $
equal to some constant
which in principle depends on the
Rabi frequency $F$, interaction strength $u$ and the fluctuator
relaxation rates. 

\begin{figure}[h]
 \centerline{
\includegraphics[width=8cm]{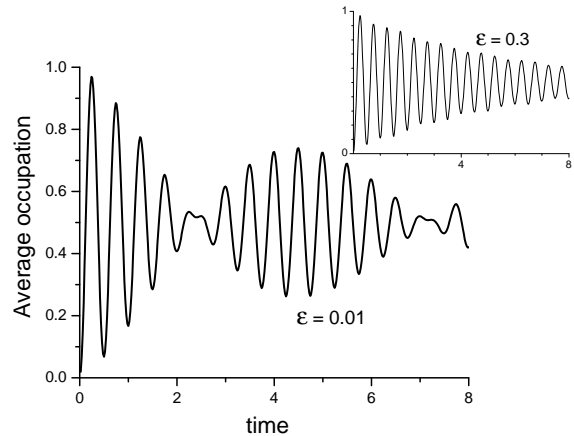}  
 }
 \caption{The qubit upper level occupation, $n_{\ua \da} +
   n_{\ua  \ua}$, as a function of time measured in Rabi periods,
   $F^{-1}$. The parameters are: $\gamma =0.03F$, $\eta=0.1$, ${\cal
   E}=0$ and $  n_{\ua \ua}/n_{\ua \da}=0.43 $. Values of the 
   fluctuator's deviation from the resonance, $\epsilon$, are shown
   near the curves.   \label{fig:res-nores}}
 \end{figure}

We have calculated  the probability to find the qubit in its upper
state,  $n_{\ua \ua}(t) + n_{\ua \da}(t)$, from
Eq.~(\ref{eq:07}) and plotted it as a function of time in
Fig.~\ref{fig:res-nores}.

The qubit is assumed to be in resonance with the pumping field,
${\cal E}=0$. We observe clear Rabi oscillations with frequency $F$
and a decay in time due to finite relaxation, $ \gamma =0.03 F$
The oscillations are perturbed by the fluctuator coupled to the qubit
with the strength $\eta=0.1$. Two curves correspond to different
shifts of the fluctuator from the resonance, $\epsilon = \pm 0.3$ (these two
curves are practically the same) and
$\epsilon=0.01$. 
One can see that close to the
resonance there are beats in the Rabi oscillations. 
This should be expected from Fig. \ref{fig:01} since at the resonance
both transitions 
participate. Also, for times shorter than
the inverse beating frequency or larger damping the resonance effect
 amounts to a 
suppression of the Rabi oscillations. 
It is clear that the closer the fluctuator is to 
resonance the stronger it suppresses Rabi oscillations. This fact
confirms that suppression of Rabi oscillations observed in
Ref.~\onlinecite{Martinis} can indeed be induced by coupling to a
fluctuator close to resonance. 
Large damping, certainly, makes the beats less visible.

Experimentally, the qubit state is probed spectroscopically by
inducing transitions from the upper state to some high-energy  state
having small lifetime due to tunneling.~\cite{Martinis} 
 One can start from the assumption that 
 the transition probability is independent of the fluctuator
state. 
This is why we present the results for the sum  $n_{\ua \ua}+n_{\ua \da}$ 
in Fig. \ref{fig:res-nores}, see also
Ref.~\onlinecite{Loss}. 

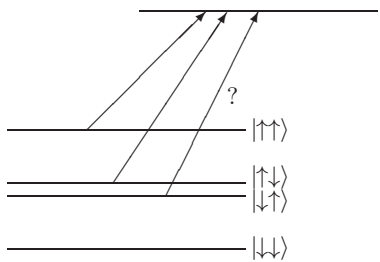
\begin{figure}
\begin{picture}(150,100)(0,0)
\put(5,5){\line(1,0){90}}
\put(5,25){\line(1,0){90}}
\put(5,30){\line(1,0){90}}
\put(5,50){\line(1,0){90}}
\put(55,95){\line(1,0){90}}
\put(97,3){$|\!\!\da\da\rangle$}
\put(97,21){$|\!\!\da\ua\rangle$}
\put(97,30){$|\!\!\ua\da\rangle$}
\put(97,48){$|\!\!\ua\ua\rangle$}
\put(35,50){\vector(1,1){45}}
\put(45,30){\vector(2,3){43}}
\put(65,25){\vector(1,2){35}}
\put(88,60){?}
\end{picture}
\caption{Energy levels and transitions.\label{levels}}
\end{figure}

In reality, the transition probabilities from the states $|\! \!\ua
\ua\rangle$ and $|\! \! \ua \da\rangle$ can significantly differ. Moreover, 
transitions from the qubit down states  $|\! \!\da
\ua\rangle$ and $|\! \! \da \da\rangle$ can be induces. For example, 
if the qubit is resonant with the fluctuator, the
levels $|\!\!\ua\da\rangle$ and $|\!\!\da\ua\rangle$ will be close in
energy, while the level $|\!\!\ua\ua\rangle$ will be removed from this close 
pair (see Fig.~\ref{levels}).

It is then hard to expect that the driving force
 is resonant
with transitions from the levels $|\!\!\ua\da\rangle$ and $|\!\!\ua\ua\rangle$
without also being coupled to the $|\!\!\da\ua\rangle$ state. 
Different response of the different levels to the measurement pulse can 
significantly change the shape of the curve.
\begin{figure}[ht]
\centerline{
\includegraphics[width=6cm]{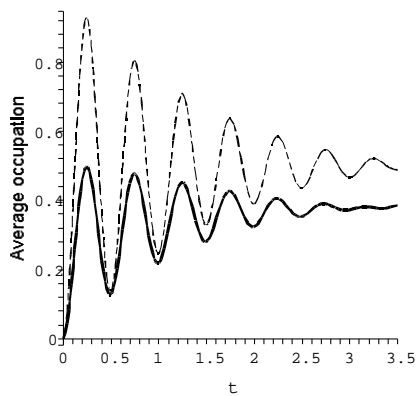}
}
\caption{Solid curve - weighted average of the qubit upper state
  population, $ 0.5n_{\ua \da} +  n_{\ua  \ua}$, for
  $\epsilon=0.01$. Other parameters are the same as in
  Fig.~\protect{\ref{fig:res-nores}} . For
  comparison the sum $n_{\ua \da} + n_{\ua  \ua}$ is shown by dashed
  curve.\label{fig:superposition}} 
\end{figure}
To illustrate this point
we plot in Fig.~\ref{fig:superposition} a weighted average, $
0.5n_{\ua \da} + n_{\ua \ua}$, assuming that the escape 
from the state $|\! \! \ua \ua\rangle$ is twice as probable as the escape 
 from the state
$|\! \! \ua \da\rangle$(solid curve).
Interestingly, this curve is much
more similar to the experimentally observed one~\cite{Martinis} than the simple
sum shown for comparison (dashed line). Namely, the plotted quantity
oscillates never exceeding 1/2 and tends to $\approx 1/2$ at large
time. As far as we understand it, it is  
unclear what combination of the occupation 
numbers of the  states is actually probed in the experiment. In 
the second set of experiments\cite{Martinis1} a different readout scheme 
was applied, but no Rabi oscillation traces where presented.


In conclusion, we develop a general scheme of analytical analysis of
time-evolution of the density matrix of a multi-level quantum system
coupled to a heat bath. In this scheme the dissipation is taken into
account phenomenologically, while the quantum dynamics in the presence
of arbitrarily strong driving force is evaluated explicitly. We
applied this scheme to the system of a qubit coupled to a fluctuator
which probably was experimentally realized in Ref~\onlinecite{Martinis}. We
demonstrated that if the fluctuator is close to resonance with the
qubit, the Rabi oscillations of the qubit are suppressed at short
times and demonstrate beatings when the damping is weak enough. We
also found that if the read-out signal depends on the state of the
fluctuator, the visibility of the Rabi oscillations can be substantially
reduced. This effect can naturally explain the experimental results of
Ref~\onlinecite{Martinis}.


\acknowledgments
This research is supported by the Norwegian Research Council,
FUNMAT@UiO and the US DOE Office of
Science under contract No. W-31-109-ENG-38, ARO/ARDA (DAAD19-02--1-0039)
DARPA under QuIST program.


\begin{thebibliography}{99}
 \bibitem{Martinis} R. W. Simmonds, K. M. Lang, D. A. Hite, S. Nam,
 D. P. Pappas, and John M. Martinis. \prl {\bf 93}, 077003 (2004).
\bibitem{Martinis1}
K. B. Cooper, M. Steffen, R. McDermott, R. W. Simmonds, Seongshik
 Oh, D. A. Hite, D. P. Pappas, John M. Martinis, cond-mat/0405710.
\bibitem{Nakamura} Y. Nakamura, Y. A. Pashkin, and J. S. Tsai, Nature
  {\bf 398}, 786 (1999).
\bibitem{Martinis2} J. M. Martinis, S. Nam, J. Aumentado, C. Urbina,
  \prl {\bf 89}, 117901 (2002).
\bibitem{Vion} D. Vion, A. Aassime, A. Cottet, P. Joyez, H. Pothier,
  C. Urbina, D. Esteve, and M. H. Devoret, Science {\bf 296}, 886
  (2002).
\bibitem{Yu} Y. Yu, S. Han, X. Chu, S. Chu, and Z. Wang, Science {\bf 296}, 889
  (2002). 
\bibitem{Nakamura2} I. Chiorescu, Y. Nakamura, C. J. P. M. Harmans,
  and J. E. Mooij, Science {\bf 299}, 1869 (2003). 
     
\bibitem{Loss} F. Meier and D. Loss, cond-mat/0408594.
\bibitem{Ku} L.-C. Ku and C. C. Yu, cond-mat/0409006.
\bibitem{book} C. P. Slichter, ``Principles of Magnetic Resonance'',
  Springer Verlag, Berlin Heidelberg 1990.
\bibitem{GAS} Y. M. Galperin, B. L. Altshuler, D. V. Shantsev, in
  ``Fundamental Problems of Mesoscopic Physics'', ed. by I. V. Lerner
  \textit{et al.}, 2004 Kluwer Academic Publishers, the Netherlands,
  p. 141-165; cond-mat/0312490.
 \end{thebibliography}
\end{document}